# Metal-to-insulator transition in anatase TiO$_2$ thin films induced by growth rate modulation

T. Tachikawa[1,2,a)], M. Minohara[1], Y. Nakanishi[2], Y. Hikita[1], M. Yoshita[3], H. Akiyama[3], C. Bell[1], and H. Y. Hwang[1,4]

[1]*Stanford Institute for Materials and Energy Sciences, SLAC National Accelerator Laboratory, Menlo Park, CA 94025, USA.*

[2]*Department of Advanced Materials Science, The University of Tokyo, Kashiwa, Chiba 277-8561, Japan.*

[3]*Institute for Solid State Physics, The University of Tokyo, Kashiwa, Chiba 277-8581, Japan.*

[4]*Department of Applied Physics, Stanford University, Stanford, CA 94305, USA.*

**Abstract**

We demonstrate control of the carrier density of single phase anatase $TiO_2$ thin films by nearly two orders of magnitude by modulating the growth kinetics during pulsed laser deposition, under fixed thermodynamic conditions. The resistivity and the intensity of the photoluminescence spectra of these $TiO_2$ samples, both of which correlate with the number of oxygen vacancies, are shown to depend strongly on the growth rate. A quantitative model is used to explain the carrier density changes.

a) Author to whom correspondence should be addressed; Electronic mail : (rapid@slac.stanford.edu)

Anatase titanium dioxide (TiO$_2$) is a functional oxide employed for various electronic applications. For example, when doped with transition metal ions, anatase TiO$_2$ can be a transparent conducting electrode [1, 2] or exhibits magnetism [3]. A technical requirement essential for the incorporation of anatase TiO$_2$ thin films in devices is the ability to control the density and mobility of the free carriers originating from point defects – here oxygen vacancies. In the case of pulsed laser deposition (PLD) of thin oxide films, utilized in this work, the most common strategy for controlling defects is to optimize the thermodynamic conditions based on the defect equilibria [4]. However, in the case of anatase TiO$_2$, this strategy is hindered by the comparable formation free energy of the rutile phase [5]. This narrows the thermodynamic window available to obtain single phase anatase TiO$_2$ [6]. Therefore, an independent strategy based on growth kinetics, which varies the adatom surface migration, the adsorption/desorption dynamics, and the island nucleation density, must be applied to control the defects and hence the carrier density and mobility.

It is known that growth kinetics can be modulated by the growth rate in various systems [7-9]. Previously, the growth rate was found to strongly influence the structural quality of Cr-doped anatase TiO$_2$ thin films grown by molecular beam epitaxy [10]. In that case it was shown that the thin films grown at lower growth rate (~0.0015 nm/s) were less defective compared with those grown at higher growth rate (~0.01 nm/s). It was suggested that the layer-by-layer growth mode, realized using a slow growth rate, was the probable origin for the defect density change.

Here, using PLD to grow nominally undoped anatase TiO$_2$, we investigated in detail the effect of the growth rate on the defect density. The defect density was characterized using electrical transport and photoluminescence (PL) studies. By varying the growth rate, a metal-to-insulator transition was observed, driven by a change of nearly two orders of magnitude in the carrier density at fixed growth temperature and oxygen pressure. PL measurements independently confirmed that the carriers originate from oxygen vacancies in the films. These results can be modeled quantitatively by considering the competition between the time scales of oxygen adsorption on the growth surface and loss from the subsurface.

Anatase $TiO_2$ films were epitaxially grown on $LaAlO_3$ (001) substrates by PLD with a KrF excimer laser (wavelength = 248 nm, pulse duration = 25 ns) focused to a spot of area 0.04 cm$^2$ on a polycrystalline target. Prior to deposition, the substrates were annealed at 900 ºC under an oxygen pressure of $1 \times 10^{-6}$ Torr to obtain single terminated $AlO_2$ surfaces with a step and terrace structure [11]. The thermodynamic conditions during growth, the substrate temperature and the oxygen partial pressure, were fixed at 700 ºC and $1 \times 10^{-5}$ Torr respectively, similar to those reported previously for high quality epitaxial anatase $TiO_2$ thin films [1, 12-16]. The growth rate, defined as the total thickness (fixed at 60 nm) divided by the total deposition time, was controlled by the laser fluence and the laser repetition rate. The former varies the amount of the ablated species per laser shot, and the latter determines the temporal variation in the density of the species at the growth surface [8, 17]. Using atomic force microscopy, all film surfaces were found to have root mean square surface roughness values of less than 1 nm, and x-ray diffraction measurements confirmed the successful growth of single phase epitaxial (001)-oriented anatase TiO$_2$ thin films under all conditions within experimental resolution.

The variation of the temperature dependence of the resistivity ($\rho$-$T$) of the anatase TiO$_2$ films for various growth rates was non-monotonic, as shown in Fig. 1(a). A metal-to-insulator transition was observed as the growth rate, $r$, decreased from 18.3 × 10$^{-3}$ nm/s to 6.6 × 10$^{-3}$ nm/s, while the opposite behavior was found for 6.6 × 10$^{-3}$ nm/s $\leq r \leq 1.9 \times 10^{-3}$ nm/s. The resistivity at $T$ = 300 K, $\rho_{300\,K}$, as a function of $r$ [shown in Fig. 1(c)], peaks around a value of $r \sim$ 5-7 × 10$^{-3}$ nm/s. This peak value coincides with a minimum in the carrier density obtained from Hall measurements [Fig. 1(c)]. The maximum change in the carrier density is nearly two orders of magnitude, contrasting with a marginal variation in the electron Hall mobility [Fig. 1(b)].

In order to understand the origin of these data, it is important to note that reflection high energy diffraction was used to confirm both (1 × 4) reconstructed surfaces during growth, as well as the film thickness from the intensity oscillations. The latter showed good agreement with *ex-situ* stylus profilometer measurements indicating layer-by-layer growth for all samples [16]. Therefore this non-monotonic variation of the resistivity and carrier density cannot be explained by a growth mode change. In particular, the carrier density variation suggests a change in the number of defects, which are well known to act as dopants. In this scenario, we must consider the type of defects which change with $r$. From Hall measurements, the carriers are electrons, suggesting oxygen vacancies ($V_O^{\bullet\bullet}$) or Ti interstitial ($Ti_i^{\bullet\bullet\bullet}$) as the origin of the carriers. For the insulating samples, around the peak of $\rho_{300\,K}(r)$, the activation energy estimated from an Arrhenius plot of the $\rho$-$T$ data at relatively high temperatures was in the range 30-60 meV. These values are of similar order to the electron activation energy level for $V_O^{\bullet\bullet}$ obtained from first principles calculations, and rather small compared to that for $Ti_i^{\bullet\bullet\bullet}$ [18]. Since this energy scale is similar to the thermal energy at $T = 300$ K, the carrier density from Hall measurements should be comparable to the $V_O^{\bullet\bullet}$ density $N(V_O^{\bullet\bullet})$.

In order to obtain further insights into the origin of the defect-generated carriers, PL measurements were performed using a He-Cd laser ($\lambda = 325$ nm) excitation source at $T = 10$ K. The PL spectrum for the sample grown with $r = 18.3 \times 10^{-3}$ nm/s is shown in Fig. 2(a). The main peak due to the anatase $TiO_2$ is clearly observed around 2.2 eV, with a much smaller peak around 1.7 eV originating from $Cr^{3+}$ impurities in the $LaAlO_3$ substrate [19]. Similar data were obtained for a total of four different samples. Given the strong electron-phonon interaction of anatase $TiO_2$, the PL is expected to originate from a self-trapped exciton (STE). However, the surface state must also be considered, otherwise the full widths at half maximum of the spectra are significantly larger than the 0.4 eV expected from the reported characteristic phonon energy and the Huang-Rhys factor for PL from only the STE [20]. The PL spectra were therefore modeled as a combination of two Poisson components, $\alpha$ and $\beta$ [21], peaked at 2.2 eV and 2.4 eV as shown in Fig. 2(a) [22], corresponding to the surface state [23] and the STE coupled to $V_O^{\bullet\bullet}$ [21, 24], respectively. Notably, $Ti_i^{\bullet\bullet\bullet}$ related luminescence, expected between 1.4 eV to 2.1 eV from density functional theory calculations, is absent [18]. This is reasonable since the oxygen growth pressure is in a regime where the formation of $Ti_i^{\bullet\bullet\bullet}$ is energetically not favorable [25]. A clear scaling is found between the normalized integrated intensity of the β peak, given by $A_β/A_{total}$ with the Hall carrier density (Fig. 3). All of these points suggest that $V_O^{\bullet\bullet}$ is the dominant defect which dopes carriers in these samples.

As further confirmation, the excitation power ($P$) dependence of the PL from the STE was used to quantitatively measure the relative change of $N(V_O^{\bullet\bullet})$ with $r$ [26]. A schematic of the formation processes of the STE is shown in Fig. 2(b) [24]. Both indirect ($β_1$) and direct ($β_2$) formation processes occur, with the former including trapping by $V_O^{\bullet\bullet}$ states and re-excitation to the conduction band. As $P$ increases, the $V_O^{\bullet\bullet}$ states in the $β_1$ process are progressively filled with optically excited carriers, until a critical excitation power ($P_{crit}$) beyond which further photo-excited carriers recombine only via $β_2$ and non-radiative channels. The intensities of the β peaks, $A_β$ in Fig. 2(a) measured for different $P$ are shown in Fig. 2(c) for the case where $r = 6.6 \times 10^{-3}$ nm/s. At low $P$, $A_β$ is linearly proportional to $P$, but becomes sub-linear above $P_{crit}$, as indicated. Using the form $A_β \approx P^γ$, $γ$ is calculated by taking the derivative of a cubic spline interpolation of $A_β(P)$, and $P_{crit}$ is defined as the value of $P$ where $γ$ drops below 0.95. The excitation power dependence of $γ$ is shown in Fig. 2(d) for all four samples. The form of $P_{crit}(r)$ closely matches the carrier density-growth rate variation, as shown in Fig. 3. From these electrical transport and PL results, we can conclude that $V_O^{\bullet\bullet}$ defects are the dominant source of the generated carriers, and that the growth rate variation directly dictates $N(V_O^{\bullet\bullet})$.

In order to quantitatively understand why the oxygen defect density changes with the growth rate, we start from the fact that $N(V_O^{\bullet\bullet})$ in the ablated precursors ($C_p$) is far larger than that of the growth surface ($C_s$) and the subsurface ($C_{ss}$) local equilibrium states [27]. In this circumstance, ablated precursors which are adsorbed on the surface of the growth-front, are oxidized easily. Furthermore, first principles calculations have shown that $C_s$ is slightly smaller than $C_{ss}$, meaning that the subsurface of anatase TiO$_2$ (001) favors a more reduced state than the surface [28]. Therefore, the balance between oxidization at the surface, characterized by the time $\tau_s$, and the timescale of oxygen loss from the subsurface ($\tau_{ss}$) during growth is crucial to determine the final $N(V_O^{\bullet\bullet})$, neglecting oxygen transfer from deeper layers [15]. When the time to grow one unit cell ($\tau$) is too short to reach $C_{ss}$ and $C_s$ ($\tau \ll \tau_s < \tau_{ss}$), a relatively large $N(V_O^{\bullet\bullet})$ remains in the film. This corresponds to the case of large $r$. On the other hand, for smaller $r$ ($\tau_s < \tau_{ss} \ll \tau$), while the surface is fully oxidized, there is sufficient time for significant oxygen desorption from the subsurface, and consequently, oxygen reduced anatase TiO$_2$ is again obtained throughout the film thickness. At intermediate $r$ where $\tau_s < \tau < \tau_{ss}$, oxygen desorption from the subsurface is suppressed relative to the surface oxidation, resulting in the smallest $N(V_O^{\bullet\bullet})$. We calculated the carrier density from $N(V_O^{\bullet\bullet})$ as a function of $r$ using $\tau_s = 30$ s, $\tau_{ss} = 1050$ s, $C_p = 1.47 \times 10^{20}$ cm$^{-3}$, $C_s = 1.47 \times 10^{17}$ cm$^{-3}$ and $C_{ss} = 2.25 \times 10^{19}$ cm$^{-3}$ estimated from previous reports [15, 25, 27-30] as shown in Fig. 3 (solid line). The good agreement with the experimental results confirms that the kinetic balance of the time constants dominate the formation of $V_O^{\bullet\bullet}$ in these TiO$_2$ films during growth.

In summary, we succeeded in controlling the carrier density of anatase TiO$_2$ thin films by two orders of magnitude through growth rate control using fixed thermodynamic conditions. This variation was caused by the growth rate dependence of the number of $V_O^{\bullet\bullet}$ defects, and could be explained quantitatively by considering the balance between the adsorption and desorption processes during growth. This technique opens up the possibility of tuning the electrical and optical properties of $TiO_2$ while maintaining the anatase single crystal phase, and provides a flexible way to control crystal defects in more complex heterostructures, where changes to the thermodynamic conditions may be detrimental to other pre-deposited layers.

The authors thank M. Lippmaa for experimental support. T. T., M. M., Y. H., C. B. and H. Y. H. acknowledge support from the Department of Energy, Office of Basic Energy Sciences, Materials Sciences and Engineering Division, under contract DE-AC02-76SF00515.

**Figure Captions**

FIG. 1. (color online): (a) $\rho$-$T$ curves of anatase $TiO_2$ for various growth rates. Numbers correspond to the growth rate ($\times\ 10^{-3}$ nm/s). (b) Hall mobility, (c) carrier density and resistivity at $T = 300$ K, $\rho_{300\ K}$, as a function of the growth rate, $r$. The dashed lines are guides for the eye. The solid line gives the calculated carrier density from the coupled rate equation described in the text.

FIG. 2. (color online): (a) PL spectrum of anatase $TiO_2$ with $r = 18.3 \times 10^{-3}$ nm/s (solid line) and the decomposition of the spectrum into two Poisson curves: $\alpha$, (open dotted line), and $\beta$, (dashed, filled curve). Sum of $\alpha$ and $\beta$ is shown by a dashed line. (b) Schematic diagram of the indirect ($\beta_1$) and direct ($\beta_2$) processes that make up the $\beta$ component. $E_v$ and $E_c$ are the valence and conduction band energies. (c) Excitation power ($P$) dependence of $A_\beta$ for $r = 6.6 \times 10^{-3}$ nm/s. The dashed line shows a linear fit at relatively low $P$ and the solid line the cubic spline fit. (d) $P$ dependence of $\gamma$ in $A_\beta \approx P^\gamma$ for various $r$. Dashed lines are guides for the eye. Solid line indicates $\gamma = 0.95$ which defines $P_{crit}$. All the PL spectra were obtained at $T = 10$ K.

FIG. 3. (color online): Growth rate dependence of the carrier density, $P_{crit}$, and $A_\beta/A_{total}$. The dashed lines are guides for the eye. The solid line gives the calculated carrier density as described in the text.

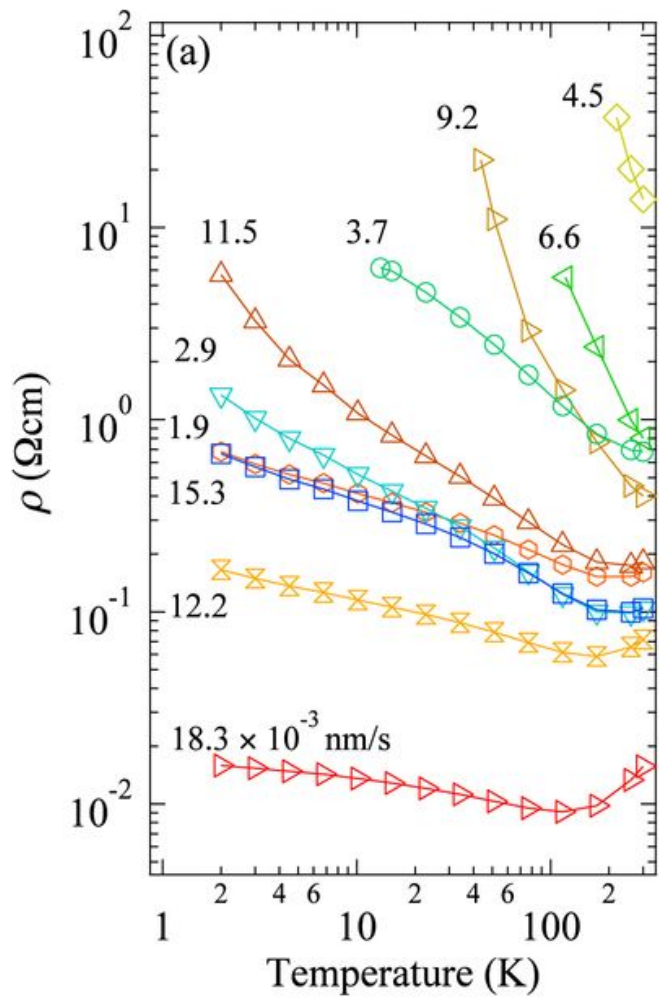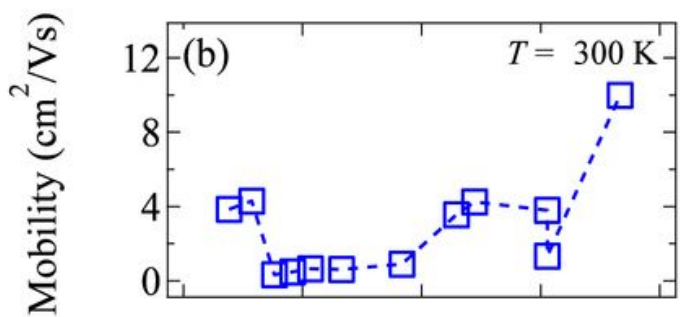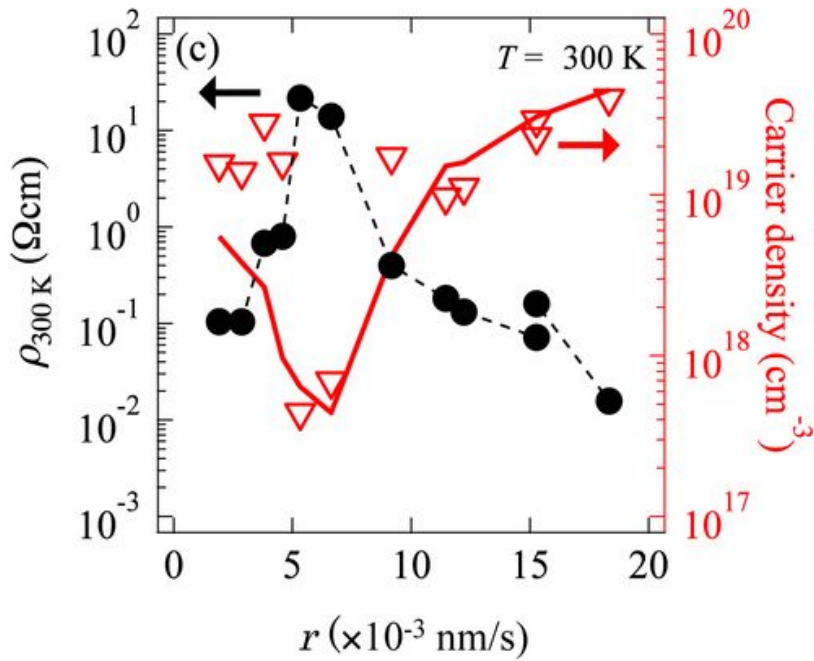

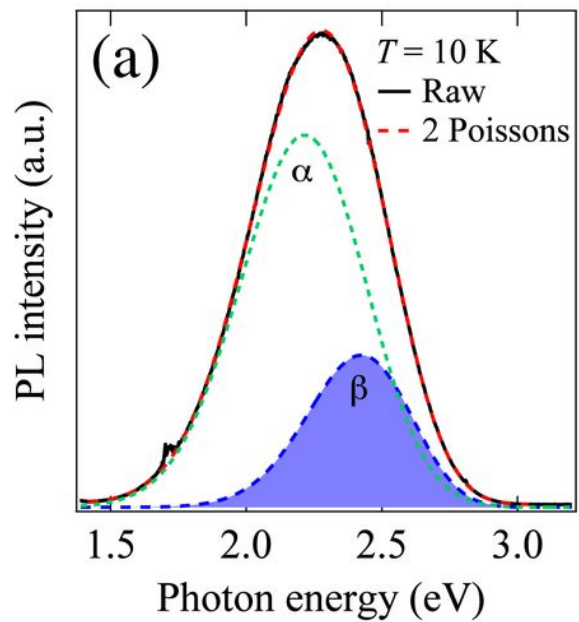
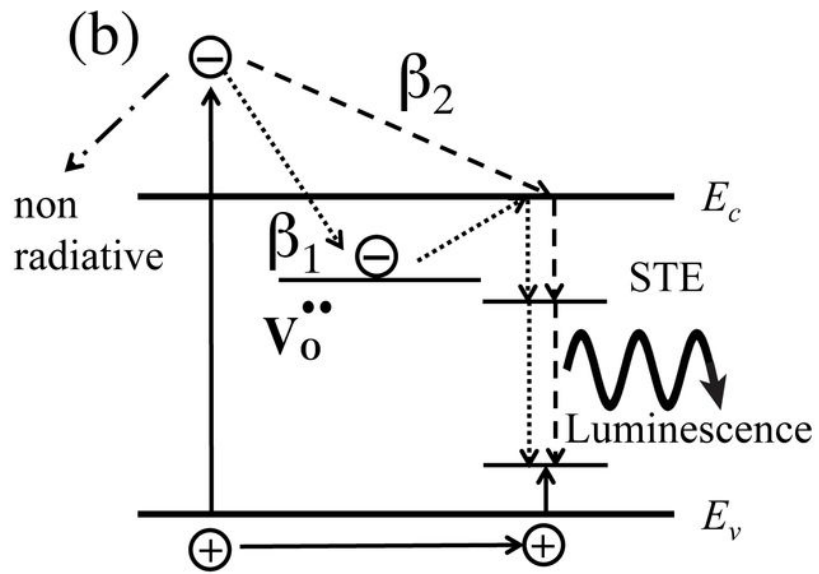
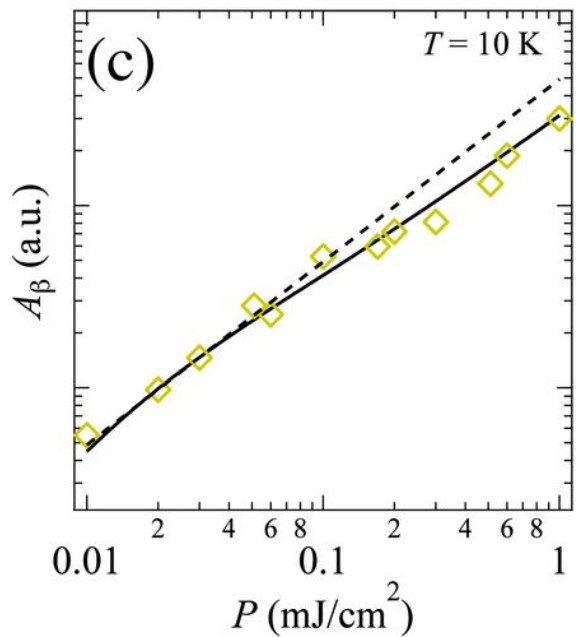
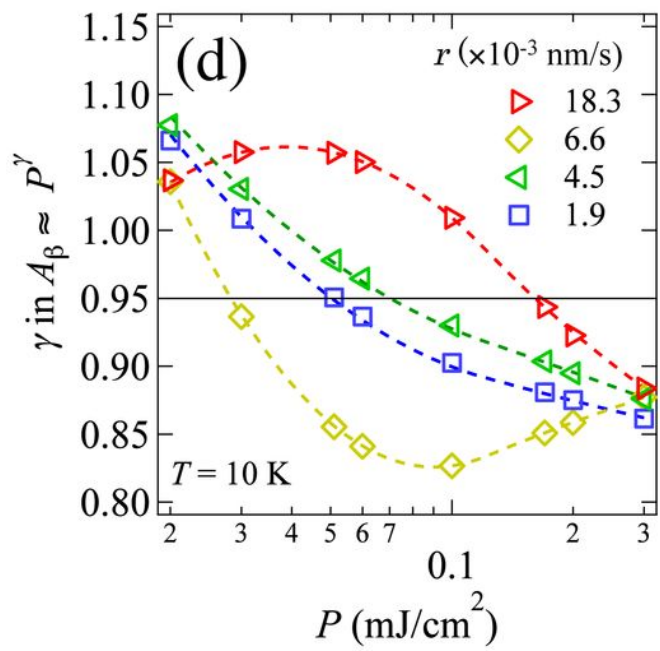

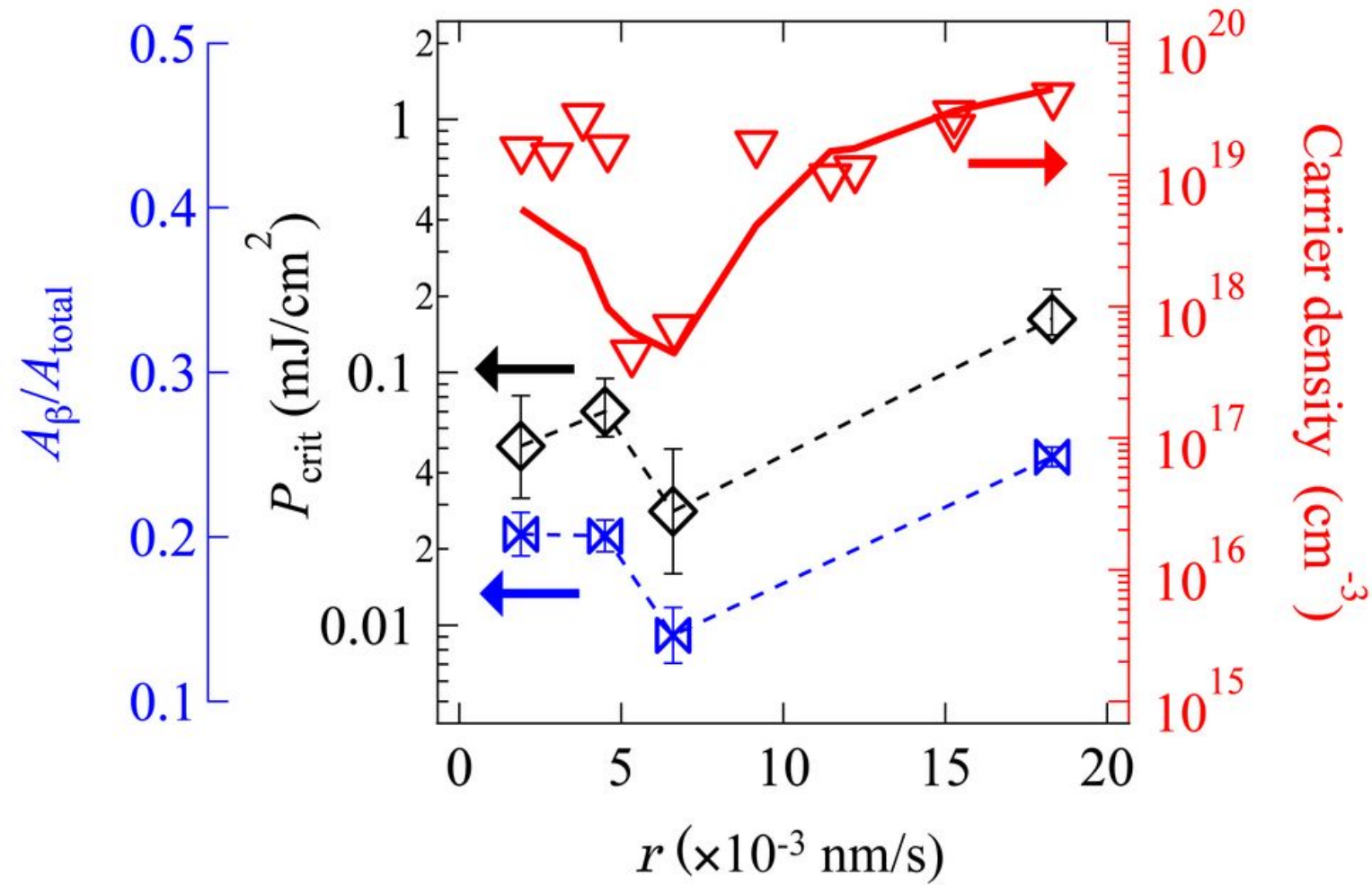